\documentclass[prb,floats,aps,preprint,superscriptaddress,showpacs]{revtex4}
\usepackage{amsmath}
\usepackage{graphicx}

\begin{document}
\title{Beyond the Dynamic Density Functional theory for steady
currents. Application to driven colloidal particles in a channel.}

\author{P. Tarazona}
\affiliation{%
Departamento de F{\'{\i}}sica Te\'orica de la Materia
Condensada, and 
Instituto de Ciencia de Materiales Nicol\'as Cabrera,
Universidad Aut\'onoma de Madrid, E-28049 Madrid, Spain}
\author{Umberto Marini Bettolo Marconi}
\affiliation{Dipartimento di Fisica, Universit\`a di Camerino and
Istituto Nazionale di Fisica della Materia,
Via Madonna delle Carceri, 62032 Camerino, Italy}

\date{\today}

\begin{abstract}
Motivated by recent studies on the dynamics of colloidal solutions
in narrow channels, we consider the steady state properties
of an assembly of non interacting particles
subject to the action of a traveling potential 
moving at a constant speed while
the solvent is modeled by  a heat bath at rest in the 
laboratory frame.
Since the description, we propose here, takes into account the
inertia of the colloidal particles it is
necessary to consider the evolution of both
positions and momenta and study 
the governing equation for the one-particle phase-space distribution.
We first derive the asymptotic form of its solutions
as an expansion in Hermite polynomials and their generic properties,
such as the force and energy balance and then
we particularize our study to the case of an inverted parabolic 
potential barrier. We  
obtain numerically the steady state density and temperature profile
and show that the expansion is rapidly convergent for large values of the
friction constant and small drifting velocities.
The present results on the one hand confirm the 
previous studies based on the 
dynamic density functional theory (DDFT) when the friction constant is 
large, on the other hand  display effects such as
the presence of a wake behind the barrier and  a strong inhomogeneity
in the temperature field which are beyond the DDFT description.

\end{abstract}
\maketitle

\section{Short introduction}
\label{introduction}
In recent years we have witnessed the emergence of a new branch of 
applied physics named
microfluidics, which is the science of designing, manufacturing
devices and processes that deal with volumes of fluid on
the order of nanoliters \cite{squires,whitesides,gad,ho,tabeling}.
Microfluidic systems have diverse and widespread potential
applications \cite{mic,labchip,pp}. Some examples of systems and processes that might employ
this technology include ink-jet printers, blood-cell-separation
equipment, biochemical assays, chemical synthesis, genetic analysis,
drug screening, electrochromatography, surface micro-machining, laser
ablation, and mechanical micro-milling. Not surprisingly, the medical
industry has shown keen interest in microfluidics technology.

Such advances in manipulating fluids \cite{cui,sullivan,bechinger} 
have recently motivated
Penna and Tarazona  \cite{pennatarazona} to consider 
a model representing a  simple device to push 
a dilute solution of colloidal particles along a narrow channel.
In particular they studied
the effect of a moving barrier on  a system of 
non interacting colloidal particles described by overdamped Langevin dynamics. 
Under the action of the  potential barrier shifting
at a constant speed, the fluid achieves a 
steady state, with density distribution and local current following
the moving barrier. These authors showed that such a steady state 
can be conveniently studied within the 
DDFT \cite{marconitarazona,archer2004,yoshimori}
formalism, since
the structure of the relevant 
equations becomes similar to that of the Euler-Lagrange equations
describing a fluid at thermodynamic equilibrium \cite{evans}.

On the other hand, the present authors  
in a recent paper~\cite{marconi2006}, hereafter referred as 
Ref. I , have considered
how the inertia of the particles may modify the DDFT picture.
They assumed that the colloidal particles 
have inertia, i.e. are governed by a second order
stochastic equation. The governing equation
for the associated phase-space distribution 
turns out to be the Kramers equation \cite{Kramers} 
and represents the evolution of both
positions and momenta of the particles.
Since such a representation is still too complex and often redundant,
the authors considered a 
contraction of such a description by rewriting the
Kramers equation in terms of the infinite
hierarchy of equations for the velocity moments
of the phase-space distribution. In ref. I the hierarchy was truncated 
systematically by means of a multiple time scale
technique, which lead to a 
self-consistent equation involving only the one-body
density. This equation is similar to 
to the DDFT equation, but contains additional terms taking into account
the presence of momentum and energy currents.
While
in ref. I we considered only transient effects, namely the decay
of initial perturbations towards the equilibrium, time
independent, state, in the present
work we illustrate how the inertial dynamics affects
the behavior of systems in situations in which a steady state 
is induced by the presence of an external time-dependent
potential.   
The results of the present paper show pronounced 
differences with respect to the DDFT study of Penna and Tarazona
and in particular display an exponential decay in the 
structure of the density profile behind the barrier which was
not predicted by the DDFT.
Moreover, we find that also the local temperature 
is non uniform throughout the system due to the heating produced by the 
barrier.

We believe that these findings are generic to 
non equilibrium systems where the
equilibration mechanism provided by the heat bath is not 
very rapid. We have shown that
when the friction is not sufficiently high the density alone is not sufficient
to characterize the steady state of the system, and additional fields
are necessary to provide a complete description.

More generally speaking,  we believe that the use of the DDFT,
is justified when the currents are of diffusive character, 
while in the cases where convective terms 
are present it is necessary to include extra terms which 
describe the transport of momentum and energy
\cite{marconimelchionna,marconicecconi}


The present paper is organized as follows. In section \ref{Theory} after 
presenting the model we give the structure of the general solution
of the Kramers equation in the region where the potential is vanishing
small. In \ref{steady}, we specialize the treatment to the case 
of steady state conditions and derive explicitly the behavior
of the  phase space distribution in the region where the
traveling potential vanishes. We also derive the relation
between the total force exerted by the barrier on the particles
and the friction due to the bath.
Finally in \ref{Numerical}  we give explicit numerical solutions 
of the Kramers equation in the case of an inverted parabolic 
barrier. We conclude the paper with a short discussion in \ref{conclusions}.

\section{Kramers equation for shifting potential barriers and its free modes}
\label{Theory}

The problem of the steady states in a fluid, under the action of a
shifting external potential, has been considered within the DDF under
several conditions and model interactions\cite{pennatarazona,3d,Pennadzu}. 
In all these
treatments the inertia of the particles did not play any role. Here we
wish to consider how the inertial effects modify that picture, and to
such a purpose we consider here the simplest case, which could
describe a dilute solution of colloidal particles dragged along a
narrow channel under the action of a moving potential barrier, modeled
by a time dependent external potential, $V_{ext}(x,t)=V_{ext}(x-ct)$,
which acts on the colloidal particles but has negligible effects on
the solvent.  To such purpose we consider here an assembly of non
interacting identical particles of mass $m$ moving in one dimension,
and described by the following stochastic dynamics~\cite{Pagnani,Cecconi}
\begin{eqnarray}
 m \frac{d^2 x}{dt^2} = -m \gamma  \frac{d x}{dt} +f_{ext}(x-ct)
+\xi(t),
\label{kramers}
\end{eqnarray}
with a bath providing the particles a friction constant $\gamma$, 
and a thermalizing noise with 
\begin{equation}
\langle \xi(t)\xi(s) \rangle  = 2 \gamma m k_B T_o \delta(t-s)\;,
\end{equation}
at temperature $T_o$. The external force associated with the traveling 
potential is  
$f_{ext}(x,t )=-\frac{d}{dx}V_{ext}(x-ct)$, and the
properties of the system can be studied by considering
the equation governing   $p(x,v,t)$, the density distribution
in phase space of a single particle. 
The associated
Kramers equation~\cite{Risken,Gardiner} reads,
\begin{eqnarray}
\frac{\partial}{\partial t}p(x,v,t)+
\Bigr [v \frac{\partial}{\partial x}+\frac{f_{ext}(x-c t)}{m}
\frac{\partial}{\partial v} \Bigr] p(x,v,t)=
\gamma \Bigr[\frac{\partial}{\partial v} v
+\frac{T_o}{m}\frac{\partial^2}{\partial v^2} \Bigr] p(x,v,t)
\label{fokker}
\end{eqnarray}
We assume that 
the shifting external potential is localized within a finite region
and vanishes outside. Therefore, far away from such a region
we should have a time independent equilibrium distribution $p_o(x,v)=\rho_o 
\exp(-v^2/(2 v_T^2))/
(\sqrt{2 \pi} v_T)$, where $\rho_o$ is the density of particles and  
$v_T=\sqrt{k_B T_o/m}$ the Gaussian width for their velocity distribution.
For a static external potential, i.e. the $c=0$ limit of \eqref{kramers},
the distribution $p(x,v,t)$ would evolve in time towards the thermal
equilibrium value $p_{eq}(x,v) =p_o(x,v) exp(-V_{ex}(x)/k_B T_o)$,
which would be reached (sooner or later) from any initial distribution
$p(x,v,0)$.
For $c\neq 0$ the continuous shift of the external potential 
implies a permanent 
perturbation of the thermal equilibrium, but still there would be a transient
evolution from any $p(x,v,0)$ to a unique stationary state $\tilde{p}(x-ct,v)$
in which the time dependence is reduced to a shift of the $x$ coordinate, to
follow the external potential $V_{ext}(x-ct)$. This steady state
is the object of the present study.
All the results presented here may be translated
to a purely static distribution in the presence of a time independent
external potential $V(x)=- f_o x + V_{ext}(x)$, with a constant slope
plus the same potential barrier which we are considering in
eq. \eqref{fokker}. The time derivative in the first term of eq.
\eqref{fokker}
vanishes, but there is an extra term proportional to $f_o$, to take
into account the constant background force added to the localized
barrier force $f_{ext}(x)$.  Away from the barrier the particles move
at constant mean velocity $v_o=f_o/(m \gamma)$, and a change of
reference framework from $v$ to $v'\equiv v-v_o$, leads to exactly the
same equation \eqref{fokker} for $p(x,v',t)$, when the barrier appears
as moving at rate $c=-v_o$. These two equivalent versions of the same
problem have been studied within the DDF formalism~\cite{pennatarazona,
Reimann,marchesoni,lowen2007},
valid for large $\gamma$.  The same exact mapping between the moving
barrier in a flat background and the static barrier in a sloped
potential would be valid for partially damped systems explored here.
 
It is convenient to introduce the following dimensionless variables:
\begin{equation}
\tau\equiv t \ v_T \ \rho_o, \qquad V\equiv\frac{v}{v_T},
\qquad X\equiv x \ \rho_o,\qquad C\equiv\frac{c}{v_T}
\label{adim1}
\end{equation}

\begin{equation}
\Gamma\equiv\frac{\gamma}{v_T \ \rho_o}, \qquad
F_{ext}(X,\tau)\equiv\frac{f_{ext}(x-ct)}{m v_T^2 \ \rho_o},\qquad
P(X,V,\tau)\equiv \frac{v_T}{\rho_o} p(x,v,t),
\label{adim2}
\end{equation}

Accordingly, Kramers' evolution equation for
the phase space distribution function
can be rewritten with the help of relations (\ref{adim1}-\ref{adim2})
as:
\begin{equation}
\frac{1}{\Gamma}\frac{\partial P(X,V,\tau)}{\partial \tau}
=L_{FP}P(X,V,\tau)
-\frac{1}{\Gamma}V \frac{\partial }{\partial X}  P(X,V,\tau)
-\frac{1}{\Gamma}F_e(X,\tau) \frac{\partial }{\partial V} P(X,V,\tau)
\label{kramers0}
\end{equation}
having introduced the ``Fokker-Planck'' operator $L_{FP}$ whose  
eigenfunctions $H_{\mu}(V)$ have the property:
\begin{equation}
L_{FP} H_{\mu}(V) \equiv \frac{\partial}{\partial V}\Bigl[
\frac{\partial }{\partial V }+V\Bigl]  H_{\mu}(V) = -\mu H_{\mu}(V),
\label{fokkerp}
\end{equation}
for $\mu=0,1,...$,and have the explicitly representation:
\begin{equation}
H_{\mu}(V)\equiv \frac{1}{\sqrt{2\pi}}
(-1)^{\mu} \frac{\partial^{\mu}}{\partial V^{\mu}} \exp(-\frac{1}{2}V^2).
\label{basis}
\end{equation}

It is convenient to
define raising and lowering operators in the eigenfunctions
series, $a_{\pm} H_{\mu}(V)= H_{\mu\pm 1}(V)$, so that 
the contributions of the damping and the external forces in the last two terms of
eq. \eqref{kramers0} may be represented through
\begin{equation}
V H_{\mu}(V)=H_{\mu+1}+ \mu H_{\mu-1}(V)\equiv (a_{+}+\mu a_{-})H_{\mu}(V),
\end{equation}
and
\begin{equation}
\frac{\partial H_{\mu}}{\partial V}=- H_{\mu+1}(V)\equiv -a_{+}H_{\mu}(V)
\end{equation}
The exact solutions of eq. (\ref{kramers0}),
in the regions where the external force vanishes 
may be written in terms of the infinite series of modes, $\mu=0,1,...$, with 
the generic form \cite{marconi2006}
\begin{equation}
P^{(\mu)}(X,V,\tau)= \exp(- \mu \Gamma \tau)
\exp\left[ - \frac{a_{+}}{\Gamma}
\frac{\partial}{\partial X} \right] \left(1+ \frac{a_{-}}{\Gamma}
\frac{\partial}{\partial X} \right)^\mu H_{\mu}(V) \phi^{(\mu)}(X,\tau).
\label{solution}
\end{equation}
The function $\phi^{(\mu)}(X,\tau)$, which fully defines the
mode $\tilde P^{(\mu)}(X,V,\tau)$
represents any solution of the diffusion equation
\begin{equation}
\frac{\partial}{\partial \tau}\phi^{(\mu)}(X,\tau)=
\frac{1}{\Gamma}\frac{\partial^2}{\partial X^2}\phi^{(\mu)}(X,\tau).
\label{diffus}
\end{equation}

From \eqref{solution} in the case $\mu=0$ we obtain explicitly
\begin{equation}
P^{(0)}(X,V,\tau)= 
H_0(V) \phi^{(0)}(X,\tau) - 
\frac{H_1(V)}{\Gamma} \frac{\partial \phi^{(0)}(X,\tau)}{\partial X} +
\frac{H_2(V)}{2! \Gamma^2} 
\frac{\partial^2 \phi^{(0)}(X,\tau)}{\partial X^2} +...,
\label{auto0}
\end{equation}
which describes a density inhomogeneity, represented by the term 
$\phi^{0}(X,\tau)$, and the associated
momentum current, the term  of order $1/\Gamma$,
kinetic energy current, the term  of order $1/\Gamma^2$,
and so on. These terms are  
{\it slaved} by the density and their shapes are
given by the successive derivatives of $\phi^{0}(X,\tau)$
with respect to $X$. 
Similarly, from \eqref{solution} the solution with $\mu=1$ has the explicit
representation
\begin{eqnarray}
P^{(1)}(X,V,\tau)= 
\exp(-\Gamma \tau) \left[ \left( H_1(V) \phi^{(1)}(X,\tau) - 
\frac{H_2(V)}{\Gamma} \frac{\partial \phi^{(1)}(X,\tau)}{\partial X} +
\frac{H_3(V)}{2! \Gamma^2} \frac{\partial^2 \phi^{(1)}(X,\tau)}{\partial X^2} 
+...\right)+
\right. \nonumber 
\\
\left. + \frac{1}{\Gamma} \left(
H_0(V) \frac{\partial \phi^{(1)}(X,\tau)}{\partial X} -
\frac{H_1(V)}{\Gamma} \frac{\partial^2 \phi^{(1)}(X,\tau)}
{\partial X^2} +...\right)
\right], 
\label{auto1}
\end{eqnarray}
where the first line in the r.h.s. has the interpretation of 
a {\it master} current
inhomogeneity  $\phi^{(1)}(X,\tau)$, which {\it slaves} higher
order moments with decreasing amplitudes ($1/\Gamma$,...),
while the second line in the r.h.s. has the same structure as
 $P^{(0)}(X,V,\tau)$
with amplitude 
$\phi^{(0)}=\Gamma^{-1} \partial_X \phi^{(1)}$,
and both terms have the fast decay of the exponential pre-factor. 
 The physical
interpretation of such a combination is that an 
initially pure current fluctuation, described by
$H_1(V) \phi_1(X,0)$  would die very fast, as $\exp(-\Gamma \tau)$, 
but leaving behind a density fluctuation proportional to 
$\Gamma^{-1} \partial_X \phi^{(1)}(X,0)$, 
which would evolve diffusively. The particular
combination in (\ref{auto1}) is such that it completely cancels that 
remnant density
fluctuations, i.e. it orthogonalizes 
$P^{(1)}(X,V,\tau)$ to $P^{(0)}(X,V,\tau)$,
and leaves a purely {\it fast} decaying form.
The generic free mode of order $\mu$, is a {\it master} term
$\phi^{(\mu)}(X,\tau) H_{\mu}(V) exp(-\Gamma \tau)$, representing a density ($\mu=0$),
current ($\mu=1$), temperature ($\mu=2$), heat ($\mu=3$), etc...,
perturbation of the equilibrium distribution $p_o(x,v)$. The {\it
master} distribution $\phi^{(\mu})(X,\tau)$, slaves the perturbation
components associated to any other $H_{\mu'}(V)$, with increasing
powers of the inverse damping $1/\Gamma$, so that whole distribution
$P^{(\mu)}(X,V,\tau)$ decays towards equilibrium with an exponential
decay time $(\mu \Gamma)^{-1}$. For time independent external
potentials, the high order modes are only visible as very short
transient states of $P(X,V,\tau)$ towards $p_{o}(x,v)$, and the in the
large damping limit, $\Gamma \gg 1$, the modes are essentially reduced
to their master component \cite{marconi2006}.  We analyze in this work
the role of these modes under the continuous shift of the external
potential, for finite values of the damping constant $\Gamma$.

\section{Steady state solution}
\label{steady}
\subsection{The steady state form of the free modes}
Now, we impose the steady state condition $P(X,V,\tau)=
\tilde{P}(X- C \tau,V)$, shifting  with time to follow
the boundary conditions in the moving potential barrier, in terms of
the variable $\tilde{X}=X- C \tau$.i We analyze first the form
of the free modes of the expansion to represent the 
solution of eq.(\ref{kramers0}) in the regions where the
external force vanishes. 
Since the steady solution has the property
\begin{equation}
\frac{\partial}{\partial \tau}
\left[ P^{(\mu)}(X,V,\tau)\right]= - C 
\frac{\partial}{\partial X}
\left[ P^{(\mu)}(X,V,\tau)\right],
\label{diff}
\end{equation}
it follows that
\begin{equation} 
\frac{\partial}{\partial \tau}
\left[  \exp(- \mu \Gamma \tau) \phi^{(\mu)}(X,\tau)\right]= - C
\exp(- \mu \Gamma \tau) 
\frac{\partial}{\partial X}
\left[  \phi^{(\mu)}(X,\tau)\right],
\label{diff2}
\end{equation}
so that 
we can transform the diffusion equation \eqref{diffus} for the {\it master} distribution
into an ordinary differential equation
\begin{equation}
\frac{\partial^2}{\partial X^2}\phi^{(\mu)}(X,\tau) +
\Gamma C \frac{\partial}{\partial X}\phi^{(\mu)}(X,\tau)-
\mu \Gamma^2 \phi^{(\mu)}(X,\tau)=0,
\label{odf}
\end{equation}
whose  solutions are proportional to
$\exp(\beta_{\pm}^{(\mu)} X)$
with 
\begin{equation}
\frac{\beta_{\pm}^{(\mu)}}{\Gamma}= \frac{- C \pm \sqrt{C^2+ 4 \mu}}{2}.
\end{equation}

Finally the product  $\exp(- \mu \Gamma \tau)\ \phi^{(\mu)}(X,\tau)$
featuring in eq. \eqref{solution} has the form consistent with
eq. \eqref{diff}:
\begin{equation}
\exp(- \mu \Gamma \tau)\ \phi^{(\mu)}(X,\tau) = 
\sum_{\zeta = \pm} A_{\zeta}^{(\mu)}
\exp\left[\beta_{\zeta}^{(\mu)} (X-C\ \tau)\right],
\label{godf}
\end{equation}
The amplitudes $A_{\pm}^{(\mu)}$ determine the contribution of each mode
in any region where the external potential vanishes. Since we assumed that
the potential barrier is restricted to a finite region around 
$\tilde{X}\equiv X-C\tau\approx 0$, we shall refer as the front region to the positive values
of $\tilde X$, whereas negative values $\tilde X$ represent the wake region.

For the first mode, $\mu=0$, the exponent $\beta_{+}^{(0)}$ vanishes, so that
$A_{+}^{(0)}=1$, to represent the only possible constant contribution to
$P(X,V,\tau)$, the equilibrium distribution 
$p(x,v,t)=\rho_o H_o(v/v_T)/ v_T$,
away from the perturbation. The second exponent for $\mu=0$, is
$\beta_{-}^{(0)}=-\Gamma C$, so that it can only contribute to
$P(X- C \tau ,V)$ in the {\it front} region of the advancing
potential barrier, with an amplitude $A_{-}^{(0)}$ to be fixed by the 
boundary condition at the advancing front of the external barrier.
On the region left behind the barrier, the amplitude
$A_{-}^{(0)}$ has to vanish, since otherwise $P^{(0)}(X- C \tau ,V)$
would diverge as $\exp(-C X)$ for $X \ll 0$. Therefore, 
substituting the solution  \eqref{godf}
with $\mu=0$ into eq. \eqref{auto0} we obtain
the structure
\begin{equation}
\tilde{P}^{(0)}(\tilde{X},V)= H_0(V) + A_{-}^{(0)} e^{-\Gamma C \tilde{X}} 
\left[ H_0(V) + C H_1(V)+
\frac{ C^2 H_2(V)}{2!}  +... \right],
\label{modo0}
\end{equation}
for $\tilde{X}=X-C\tau$ on the front side of the advancing barrier,
while behind the barrier we have the pure equilibrium structure 
$\tilde{P}^{(0)}(\tilde{X},V)= H_0(V)$, with no 
remnant {\it wake} structure. 

  The contribution proportional to $ H_0(V)$ in $ P^{(0)}(\tilde{X},V)$
has precisely the shape obtained from the analysis of 
eq.(\ref{kramers}) in the strong damping limit~\cite{pennatarazona}, 
when the particles are
always at their limit velocity and the inertial term can be neglected.
In this limit the Smoluchowski \cite{Smoluchowski} 
description of the system is sufficient,
and the solution can be written as $P(X,V,\tau)=
\rho(X,\tau) H_o(V)/\rho_o$, where $\rho(X,\tau)$ satisfies 
the following diffusion equation with drift
\begin{eqnarray}
\frac{\partial \rho(X,\tau)}{\partial \tau}=
\frac{1}{\Gamma} \frac{\partial^2 \rho(X,\tau)}{\partial X^2}-
\frac{1}{\Gamma}
\frac{\partial}{\partial X}\left( \rho(X,\tau) 
F_e(X-C\tau) \right),
\label{smolu0}
\end{eqnarray}
and the stationary solution $\rho(X-C \tau)$ for shifting potential barriers,
has the exponential front and the complete lack of wake identical to
the $H_0(V)$ contribution to (\ref{modo0}).  The only qualitative difference between the
fully damped system described by the Smoluchowski equation, 
and the $\mu=0$ mode solution, of \eqref{modo0}
is that the front density perturbation
slaves a current $C H_1(V)$, a kinetic energy increase $C^2 H_2(V)/2$,
and similar higher order terms which may be resumed to give exactly
the form
\begin{equation}
\tilde P^{(0)}(\tilde{X},V)= H_0(V) + A_{-}^{(0)} e^{-\Gamma C \tilde{X}} 
H_0(V-C),
\label{modo0b}
\end{equation}
i.e. the whole perturbation of $\tilde P^{(0)}(\tilde{X},V)$ over
the equilibrium value $ H_0(V)$ has a Maxwellian distribution of 
velocities but shifted to the reference frame of the advancing
potential barrier. 

All higher order terms are characterized by $\beta_{+}^{(\mu)}>0$ and
$\beta_{-}^{(\mu)}<0$, so that
the exponent $\beta_{-}^{(\mu)}$, has to be taken at
the front side and $\beta_{+}^{(\mu)}$ behind the 
barrier, so that there is one free amplitude $ A_{\pm}^{(\mu)}$
for each mode at each side of the barrier.
The distribution functions for these modes may also be written
in terms of the shifted eigenfunctions of the 
operator $L_{FP}$,  
$H_{\nu}(V+\beta_{\pm}^{(\mu)}/\Gamma)$. Thus, the $\mu=1$ mode
has the form
\begin{equation}
\tilde P^{(1)}(\tilde X,V)= 
 A_{\pm}^{(1)} e^{\beta_{\pm}^{(1)} \tilde X}  \left[ 
H_1\left(V+ \frac{\beta_{\pm}^{(1)}}{\Gamma}\right)+\frac{\beta_{\pm}^{(1)}}{\Gamma}
H_0\left(V+\frac{\beta_{\pm}^{(1)}}{\Gamma}\right) \right],
\label{modo1}
\end{equation}
whereas  for the $\mu=2$ mode we find:
\begin{equation}
\tilde P^{(2)}(\tilde X,V)= 
 A_{\pm}^{(2)} e^{\beta_{\pm}^{(2)} \tilde X}  \left[ 
H_2\left(V+\frac{\beta_{\pm}^{(2)}}{\Gamma}\right)+2\frac{\beta_{\pm}^{(2)}}{\Gamma}
H_1\left(V+\frac{\beta_{\pm}^{(2)}}{\Gamma}\right)
+(\frac{\beta_{\pm}^{(2)}}{\Gamma})^2 H_0\left(V+\frac{\beta_{\pm}^{(2)}}{\Gamma}\right)
 \right],
\label{modo2}
\end{equation}
and the generic structure of the $\mu$ mode is
\begin{eqnarray}
\tilde P^{(\mu)}(\tilde X,V)= 
 A_{\pm}^{(\mu)} e^{\beta_{\pm}^{(\mu)} \tilde X} \left(1+
\frac{ \beta_{\pm}^{(\mu)} a_{-}}{\Gamma}\right)^\mu 
H_\mu\left(V+\frac{\beta_{\pm}^{(\mu)}}{\Gamma}\right).
\label{modomu}
\end{eqnarray}
Notice that  all contributions $P^{(\mu)}(\tilde X,V)$ for $\mu>0$ 
decay exponentially away from the barrier.

The inclusion of higher order terms creates a wake density fluctuation
structure, with exponential decays $exp(\beta_{+}^{(\mu)} \tilde X)$, which
have $\beta_{+}^{(\mu)} \approx \sqrt \mu$ for $C\ll1$ and
$\beta_{+}^{(\mu)} \sim \mu/C \ll \mu$ for $C\gg \mu$. The front density
structure contains several exponential decays $exp(\beta_{-}^{(\mu)}
\tilde X)$, with $\beta_{-}/\Gamma \approx -\sqrt \mu$ for $C\ll 1$ and 
$\beta_{+}^{(\mu)}/\Gamma
\approx -C$ for $C\gg \mu$. 
Both at the front and the wake
regions, the density fluctuations go together with fluctuations in the
velocity distribution, which may be described as shifted equilibrium
distributions, $H_0(V+\beta_{\pm}^{(\nu)})$, shifted current
distributions $H_1(V+\beta_{\pm}^{(\nu)})$, etc...
 The front region is broad if the damping is weak
and the barrier velocity small, because the restoring force is proportional
to the velocity of the colloidal particles with respect to the quiescent
solvent.
The velocity distribution changes in front of the barrier
and develops secondary peaks at $V=C$, $V=-\beta_{-}^{(1)}/\Gamma$,
$V=-\beta_{-}^{(2)}/\Gamma$, etc.

%
\subsection{Generic properties of the steady state produced by 
a shifting barrier}
We consider now the generic solution of eq.  \eqref{kramers0},
including the regions inside the moving barrier, where we have to
include the force term. The steady state condition $P(X,V,\tau)=P(X-C
\tau, V)\equiv \tilde{P}(\tilde{X},V)$ transforms eq. \eqref{kramers0}
into
\begin{equation}
(V-C)\frac{\partial \tilde P(\tilde X,V)}{\partial \tilde X} = \Gamma
L_{FP}\tilde P(\tilde X,V) -F_e(\tilde X) \frac{\partial \tilde
P(\tilde X,V) }{\partial V},
\label{kramers2}
\end{equation}
The general solution of this equation may be represented as
\begin{equation}
\tilde{P}(\tilde X,V)=\sum_{\nu=0}^\infty \Phi_\nu(\tilde X) H_\nu(V),
\label{expansion}
\end{equation}
with generic functions $\Phi_{\nu}(\tilde{X})$, to be determined from
the projections of eq. \eqref{kramers2} on each of the FP
eigenfunctions $H_{\nu}(V)$.  The projections for $\nu=0$ and $\nu=1$
give
\begin{equation}
\frac{\partial \Phi_{1}(\tilde{X})}{\partial \tilde{X}}- C
\frac{\partial \Phi_{0}(\tilde{X})}{\partial \tilde{X}}=0,
\label{proj0}
\end{equation} 
and
\begin{equation}
2 \frac{\partial \Phi_{2}(\tilde{X})}{\partial \tilde{X}}- C
\frac{\partial \Phi_{1}(\tilde{X})}{\partial \tilde{X}}+
\frac{\partial \Phi_{0}(\tilde{X})}{\partial \tilde{X}}=
F_{e}(\tilde{X}) \Phi_{0}(\tilde{X})- \Gamma \Phi_{1}(\tilde{X}).
\label{proj1}
\end{equation}
The general form for any $\nu\geq 1$ is
\begin{equation}
(\nu+1) \frac{\partial \Phi_{\nu+1}(\tilde{X})}{\partial \tilde{X}}- C
\frac{\partial \Phi_{\nu}(\tilde{X})}{\partial \tilde{X}}+
\frac{\partial \Phi_{\nu-1}(\tilde{X})}{\partial \tilde{X}}=
F_{e}(\tilde{X}) \Phi_{\nu-1}(\tilde{X})- \nu \Gamma
\Phi_{\nu}(\tilde{X}).
\label{projnu}
\end{equation}
In absence of the force term $F_e(X)$, the general solution of this
(infinite) set of coupled ordinary linear differential equations may
be written in terms of the free modes \eqref{modomu}, with arbitrary
amplitudes $ A_+^{(\mu)}$ at the back side, and $A_-^{(\mu)}$ at the
front side of the moving barrier.

 The structure of eq. \eqref{proj0}
is independent of $F_{e}(\tilde{X})$, and it represents the 
continuity equation, relating the mass density 
$\rho(\tilde{X})\equiv \rho_o \Phi_{0}(\tilde{X})$ to the current density 
$j(\tilde{X})\equiv \rho_o \Phi_{1}(\tilde{X})$, to keep the mass balance 
under a steady flow,
\begin{equation}
\frac{\partial j(X,\tau)}{\partial X}=
C  \frac{\partial \rho(X-C \tau )}{\partial X}. 
\label{continuity}
\end{equation}

The integration of \eqref{proj0} from the boundary conditions
$\Phi_{0}(\tilde{X})=1$ and $\Phi_{1}(\tilde{X})=0$, far away from the
moving barrier, gives
\begin{equation}
\Phi_{1}(\tilde{X})= C (\Phi_{0}(\tilde{X})-1),
\label{curr}
\end{equation}
i.e. any positive excess $\Phi_{0}(\tilde{X})-1 \geq 0$ in the
distribution of particles near the moving barrier is associated to a
current $j(\tilde{X})=C (\rho(\tilde{X})- \rho_o)$ following the
barrier shift.  The regions with $\Phi_{0}(\tilde{X})\leq 1$ imply a
depletion of the density, and a counter-current with opposite sign to
the barrier displacement. In the strong damping limit \cite{pennatarazona} 
such a
depletion and counter-current were limited to the interior of the
potential barrier, since there was no wake left behind it. The
inertial effects here included open the possibility of such wake, so
that we may find regions outside of the moving barrier where the mean
velocity
$\langle V\rangle= \Phi_{1}(\tilde{X})/\Phi_{0}(\tilde{X})=
C[1-1/\Phi_{0}(\tilde{X})]$ has the sign opposite to $C$.

Eq. \eqref{proj1}, from
the projection of eq. \eqref{kramers2} on $H_{1}(V)$, 
represents the local balance of momentum. If we integrate
it across the whole inhomogeneity, from far from the rear to 
far from the  front of the moving potential barrier, the 
integrals of all the derivatives vanish, and we get
that the total force $F_{T}$, 
produced by the barrier on the particles balances the 
friction force created by the bath on the total current 
\begin{equation} 
F_{T}\equiv \int_{-\infty}^{\infty} 
d\tilde{X}’ F_{e}(\tilde{X}’) \Phi_{0}(\tilde{X}’)=
\Gamma \int_{-\infty}^{\infty} d\tilde{X}’ 
\Phi_{1}(\tilde{X}’),
\label{force}
\end{equation}
i.e. it gives the global force balance in the system. Notice that
only the region of potential barrier
contributes to the first integral in the left hand side, while
the entire volume contributes to the right hand side.

The integration of \eqref{proj1} from $\Phi_{0}(X)=1$, $\Phi_{1}(X)=0$, and
$\Phi_{2}(X)=0$, at any point far from the barrier gives     
the local excess of kinetic energy at any point,
\begin{equation}
\Phi_{2}(\tilde{X})=\frac{1}{2}\left[C \Phi_{1}(\tilde{X}) -
\Phi_{0}(\tilde{X})-1 + \int_{-\infty}^{\tilde{X}} d\tilde{X}’
\left(F_{e}(\tilde{X}’) \Phi_{0}(\tilde{X}’)- \Gamma
\Phi_{1}(\tilde{X}’)\right)\right].
\label{temp}
\end{equation}
Therefore, once we have the particle distribution
$\Phi_{0}(\tilde{X})$, we may get the mean velocity of the particles
$\langle V\rangle =\Phi_{1}(\tilde{X})/\Phi_{0}(\tilde{X})$ from
\eqref{curr}, and their local temperature relative to that of the
bath, $T(\tilde{X})/T_o=1+\Phi_{2}(\tilde{X})/\Phi_{0}(\tilde{X})$
from \eqref{temp}.

Similarly, the equation for $\nu=2$ in the series \eqref{projnu}
corresponds to the energy balance. Its integration from a point far
behind barrier to an arbitrary point $\tilde{X}$ gives direct access
to the heat current $\Phi_{3}(\tilde{X})$, while its integral across
the whole inhomogeneity gives the total power transferred from the
barrier to the particles
\begin{equation}
{\cal W}\equiv \int_{-\infty}^{\infty} d\tilde{X}’ F_{e}(\tilde{X}’)
\Phi_{1}(\tilde{X}’)= 2 \Gamma \int_{-\infty}^{\infty} d\tilde{X}’
\Phi_{2}(\tilde{X}’),
\label{energy}
\end{equation}
where the last integral has to be interpreted as the total heat 
dissipated by the particles due to the local temperature difference
over the bath, $\Phi_{2}=\Phi_{0}(\tilde{X}) (T(\tilde{X})/T_o-1)$.
Notice that the steady state conditions, and the fact that
the potential energy vanishes both at the front and at the rear
of the moving barrier, gives a direct relation, ${\cal W}=C F_T$,
between the power and the force. Through eqs. \eqref{curr}, \eqref{force} and \eqref{energy}
we get also a relationship between the total excess of 
particles and the excess kinetic energy. Written in terms of the  
original variables,
\begin{equation}
\int dx \rho(x) (T(x)-T_o) =\frac {m c^2}{2}
\int dx (\rho(x)-\rho_o),
\label{balance}
\end{equation}
this should be a generic property of the steady state
distributions, independent of the damping $\Gamma$.

\subsection{Expansion in terms of the steady free modes}
 The above expressions for $\Phi_{1}(\tilde{X})$,
$\Phi_{2}(\tilde{X})$,.., given in term of $\Phi_{0}(\tilde{X})$ can
only be used after the whole set of ordinary differential
eqs. \eqref{proj0}-\eqref{projnu} are solved. That requires either a
resumation of all the terms, as done for the free modes in
eq. \eqref{modo0b}, or some truncation scheme to perform a numerical
integration for the regions with $F_{e}(\tilde{X})\neq 0$.  Unless the
force is very weak everywhere, a direct truncation scheme of the
expansion in eq. \eqref{expansion}, e.g. taking
$\Phi_{3}(\tilde{X})=0$, and solving the first three equations to get
$\Phi_{0}(\tilde{X})$, $\Phi_{1}(\tilde{X})$, and
$\Phi_{2}(\tilde{X})$, leads to unphysical results, strongly dependent
on the order of the truncation. On the contrary, we have found very good
convergence, at least for any $\Gamma \geq 1$, using a finite
parametrization of $\tilde{P}(\tilde{X},V)$ based on the natural modes
for the free particles. We fix the number $\mu_{max}$ of such modes to
be used in the front and in the wake regions, so that the solution,
$\tilde{P}(\tilde{X},V)$, 
is described by $\mu_{max}+1$ constants
$A_{-}^{(\mu)}$ at the first region, and $\mu_{max}$ constants
$A_{+}^{(\mu)}$ at the second region, besides the trivial contribution
$A_{+}^{(0)}=1$.
Within the barrier region we use $2 \mu_{max}+1$
independent functions, $\psi_{\pm}^{(\mu)}(\tilde{X})$ to parametrize
$P(\tilde{X},V)$ as
\begin{eqnarray}
\tilde P(\tilde X,V)= \sum_{\mu=0}^{\mu_{max}} \sum_{\zeta=\pm}
 \psi_{\zeta}^{(\mu)}(\tilde{X}) \left(1+
\frac{ \beta_{\zeta}^{(\mu)} a_{-}}{\Gamma}\right)^\mu 
H_\mu\left(V+\frac{\beta_{\zeta}^{(\mu)}}{\Gamma}\right).
\label{inbarrier}
\end{eqnarray}
Therefore, each term $\Phi_{\nu}$ in the expansion \eqref{expansion}
is expressed as a linear combination of the 
functions $\psi_{\pm}^{(\mu)}(\tilde{X})$ to be determined by means of 
eqs. \eqref{proj0} -\eqref{projnu},for all values $\nu \leq 2 \mu_{max}+1$.

  The simplest parametrization within this scheme corresponds to
include only the $\mu=0$ mode, with $\beta_{+}^{(0)}=0$ and
$\beta_{-}^{(0)}=-C \Gamma$, so that
\begin{equation}
P(\tilde{X},V)= \psi_{+}^{(0)}(\tilde{X}) H_0(V) +
\psi_{-}^{(0)}(\tilde{X}) H_0(V-C).
\label{modo0bis}
\end{equation}
Hence, all the terms in expansion \eqref{expansion} are given in terms
of these two functions,
\begin{equation}
\Phi_0(\tilde{X})=\psi_{+}^{(0)}(\tilde{X})+\psi_{-}^{(0)}(\tilde{X}),
\hspace{1cm} \Phi_1(\tilde{X})= C \psi_{-}^{(0)}(\tilde{X}),
\hspace{1cm} \Phi_2(\tilde{X})= \frac{C^2}{2!}
\psi_{-}^{(0)}(\tilde{X}), \hbox{ etc...,}
\label{modo0bb}
\end{equation}

The projections of eq. \eqref{kramers2} on the first two FP
eigenfunctions are enough to determine $\psi_{+}^{(0)}(\tilde{X})$ and
$\psi_{-}^{(0)}(\tilde{X})$.  From eq. \eqref{proj0} we get that
$\psi_{+}^{(0)}(\tilde{X})$ has to be constant all over the system,
both inside and outside the potential barrier, therefore it is fixed
by the asymptotic value $\psi_{+}^{(0)}(\tilde{X})=1$, and we may use
the particle distribution
$\Phi_0(\tilde{X})=1+\psi_{-}^{(0)}(\tilde{X})$  as the only free
functional variable. Regarding now the projection on $H_{1}(V)$, we
get that the contributions from the derivatives of
$\Phi_{1}(\tilde{X})$ and $\Phi_{2}(\tilde{X})$ on the left hand side
of eq. \eqref{proj1} cancel each other, so that
\begin{equation}
\frac{\partial \Phi_{0}(\tilde{X})}{\partial \tilde{X}}=
F_{e}(\tilde{X}) \Phi_{0}(\tilde{X})- C \Gamma
(\Phi_{0}(\tilde{X})-1).
\label{proj1_DDF}
\end{equation}
which is exactly the DDF equation obtained and solved by Penna and
Tarazona \cite{pennatarazona} from the integration of \eqref{smolu0}.  Notice
that this simplest parametric description of $P(\tilde{X},V)$ is
therefore consistent with respect to the mass and momentum balances,
but it has not the flexibility to recover the equivalent balances of
energy ($\nu=2$), heat current ($\nu=3$), etc..., required by 
eqs. \eqref{projnu}. A direct substitution of
eqs. \eqref{modo0bb} into eqs. \eqref{projnu} shows that the local
balance for $\nu \geq 2$ fails by a term $C^{\nu-1}
F_{e}(\tilde{X})/(\nu-1)!$, at each $\nu\geq 2$. Such a  failure is
less important for low shifting rate, $C\ll 1$, and for modes $\nu \gg
C$.  Also, the global balance represented by
eqs. \eqref{force}-\eqref{energy} would be kept at any order
$\nu$, since the total integral of $F_e(\tilde{X})$ has to vanish.
The inertial effects appear to recover the local balances missed by
the DDF approximation, and we may include them in a systematic way
including in $\tilde{P}(\tilde{X},V)$ the contributions of the higher
order free modes. That enlarges the set of free functions
$\psi_{\pm}^{(\mu)}(\tilde{X})$, and allow the solution of eq.
\eqref{kramers2} up to higher order eigenfunctions of the FP operator.

\section{Numerical results for a parabolic potential barrier}
\label{Numerical}
As an application we study
a parabolic potential barrier of the form, 
$U(\tilde{X})=\kappa (1-\tilde{X}^2)/2$
creating a linear force $F_e(\tilde X)=
\kappa \tilde X$, restricted to the interval $-1\le \tilde{X} \le 1$.  
As we consider only the steady case we have to solve eqs. \eqref{projnu}
within the barrier, and to find the solutions matching 
with the physical solutions eq.\eqref{modomu} at the front 
($\tilde{X}\geq 1$) and at the wake ($\tilde{X}\leq -1$).
The matching of $\tilde{P}(\tilde X,V)$ inside and outside the barrier
is achieved by requiring that $\tilde{P}(\tilde X,V)$ in
eq. (\ref{kramers2}) has to be continuous at $\tilde X=\pm 1$, but with a
discontinuous first derivative with respect to $\tilde X$, to match the
discontinuity in $F_e(\tilde X)$, i.e.:
\begin{equation}
(V-C) \left( \left. \frac{\partial \tilde P(\tilde X,V)}{\partial
\tilde X}\right|_{\tilde X=1+\epsilon}- \left. \frac{\partial
\tilde{P}(\tilde X ,V)}{\partial \tilde X}\right|_{\tilde
X=1-\epsilon}\right)= \kappa \left. \frac{\partial \tilde{P}(\tilde
X,V)}{\partial V}\right|_{\tilde X=1},
\end{equation}
and a similar condition at $\tilde X=-1$. Most of the results presented
here have been obtained with $\mu_{max}=4$, i.e. with nine independent
functions $\psi_{\pm}^{(\mu)}(\tilde{X})$, besides the trivial 
$\psi_{+}^{(0)}=1$,
to ensure the correct projection of eq. \eqref{kramers2} up to order
$H_{9}(V)$.  Nearly identical results are obtained with $\mu_{max}=3$,
and even with $\mu_{max}=2$ for $\Gamma\geq 2$. However, qualitative
differences appear with respect to the DDF result ($\mu_{max}=0$),
unless we have both large damping $\Gamma$ and a low shifting rate $C$
for the barrier. In any case we have to deal with a set of linear
differential equations, with a shooting boundary problem, to get the
physical match with the free modes, so that $A_{+}^{(\mu)}=0$ at
$\tilde{X}=1$, and $A_{-}^{(\mu)}=0$ at $\tilde{X}=-1$

In Fig. \ref{fig:1} we present results for a high potential barrier,
$\kappa=10$, moving with respect to the bath at a relatively low
velocity, $C=0.2$.  For large damping the system is in the
“strong drift limit” \cite{pennatarazona}.  The density distribution is
strongly depleted within the barrier, while the density in the front
region grows to a large value, more than sixty times the asymptotic
density in this case, so that there are enough particles going over
the barrier to keep the stationary state. We observe that for $\Gamma
\geq 1$, the inertial effects have little influence in the structure
of the front.  When $\Phi_{0}(\tilde{X})$ is rescaled in terms of
$\Gamma (\tilde{X}-1)$, as in Fig. \ref{fig:2}(a), the curves collapse
into a single large $\Gamma$ limit. This is consistent with the fact,
that the velocity distribution at the front region is dominated by the
shifted Maxwellian form (\ref{modo0bis}).  The effect of reducing
$\Gamma$ below the value $1$ renders smaller the amplitude of the
exponential contribution in the formula $\Phi_0(X)=1+A \exp(-C \Gamma
X)$.  Nevertheless, for the lowest value of $\Gamma$ presented in that
figure the expansion in modes is still far from convergence for
$\mu_{\max}=4$.

In Fig.\ref{fig:2}(b) we present the structure of the wake 
by rescaling the distance from the left edge of the barrier by the
the factor  $\Gamma C$. The profile
saturates for low $\Gamma$, while is
continuously reduced as $\Gamma$ increases. This is consistent
with the no-wake prediction in the large damping limit, 
when the inertial effects are fully suppressed. 
Nevertheless, the decrease of the wake 
structure with increasing $\Gamma$ is very slow, so that the presence 
of such region, with $\rho(x)<\rho_o$ and 
hence mean velocity $\langle v\rangle=
v_{T} (1-\rho_o/\rho(x)) <0$, behind the shifting potential barrier, 
is an important qualitative effect induced by the 
inertial dynamics of the particles, and which was neglected within the DDF
analysis \cite{pennatarazona}.
 
In Fig. \ref{fig:3} we present the results for the same barrier as in Fig. 
\ref{fig:1}
but with a much larger velocity $C=2$. In the large damping limit such
a situation corresponds to a "high counter-current" regime 
\cite{pennatarazona},
in which the barrier moves too fast to produce a strong perturbation
in the density distribution.  When the bath damping parameter is
reduced, the inertial dynamics creates a strong amplification of the
front structure, which is now much more symmetric with respect to the
advancing barrier front at $\tilde{X}=1$. Roughly speaking, a half of
the particles at the front are actually within the potential barrier,
for $0\alt \tilde{X} \leq 1$. This is to be compared with the result
at low $C$ where the advancing front was mainly located at $\tilde{X}
\geq 1$. The density depletion is limited to the rear edge of the
barrier $\tilde{X}\approx -1$, and the density is never lower than
$0.5$ the asymptotic value. The structure of the wake region does not
show the strong scaling effect with $\Gamma$ observed in
Fig. \ref{fig:1} (b), for the low $C$ case.
The scaled structure of the front in terms of $\Gamma C(
\tilde{X}-1) $ is presented in Fig. \ref{fig:4}(a), and shows that
the decay of the density is still well represented by the exponential
form $A_{-} \exp(-\Gamma C \tilde{X} )$ of the zeroth order mode, but
with a $\Gamma$-dependent amplitude $A_{-}$.  The maximum amplitude of
the wake, just behind the moving barrier seems to be similar for all
the cases with large $\Gamma$, while the decay increases with
$\Gamma$. The results in terms of the scaled distance $\Gamma C (\tilde{X}+1)$
may be compared with those in Fig. \ref{fig:2}(b) for the
slowly moving barrier, and we observe that the wake extends now
further away from the barrier edge.

We turn, now, to the study of the  local rescaled temperature \cite{lopez}
obtained from eq \eqref{temp}. The
results for the $C=0.2$ case in Fig.~\ref{fig:5} indicate that the shifting
barrier produces a very strong heating of the system within the
barrier, with maximum $T(\tilde{X})\approx 25 T_o$ at the rear side of
the barrier, in the region of lower density.  At the scale of the
maximum $T(\tilde{X})$ the temperature is apparently constant at the
front side, but the inset shows that there is a sharp rise of
temperature at $\tilde{X}\leq 1$, and also we observe a kind of
“precursor plateau”, over distances of the order $\tilde{X}\approx
20/\Gamma$ from the barrier edge, and with a $\Gamma$ independent
value $T(\tilde{X})/T_o\approx 1.5$.  The width of that plateau may be
understood from the huge enhancement of the density on the front side
of the barrier, so that until the exponential decay makes $\Phi_{0}(1)
exp(-\Gamma C (\tilde{X}-1))\approx 1$, the large majority of the
particles contributing to $\Phi_{0}(\tilde{X})$ belongs to the
exponential component of that front, and the value of $T(\tilde{X})$
at the plateau would represent the temperature of the advancing front.
The structure of $T(\tilde{X})$ the wake is much narrower than at the
front, and it indicates a moderate heating within the density
depletion shown in Fig. \ref{fig:1}.

In Fig.~\ref{fig:5} we present the temperature distribution for the
high velocity case, $C=2$, described in Figs.~\ref{fig:3} and
\ref{fig:4}. The maximum temperature is $T(\tilde{X})\approx 5.5 T_o$,
and it is still located at the rear half of the barrier, associated to
the minimum density. The "precursor film" at the front is much shorter
and higher, so that only for the lowest value of $\Gamma$ may may be
interpreted as a incipient "plateau", this is consistent whit the
interpretation given above when we consider the density distributions
in Fig.~\ref{fig:3}. The most peculiar feature of $T(\tilde{X})$ at
this high value of $C$ is the appearance of local minimum between the
main maximum and the front edge of the barrier. The relative
importance of this feature increases with decreasing $\Gamma$, i.e. as
the inertial effects become more important. A possible interpretation
could be that the decrease of $T(\tilde{X})$ for $\tilde{X}\alt 1$ is
a signature of the adiabatic expansion of the ideal fluid when it
climbs the potential barrier. Therefore it should be restricted to
large $C$ and low $\Gamma$, to avoid the thermalization with the bath.

 Finally, we present in Fig.~{fig:7} the results for the total force
$F_T$, from eq.~\eqref{force}, obtained both for the low and high
shifting rates, as functions of the damping $\Gamma$. We present the
results for three different choices of the parametrization,
$\mu_{max}$ from $2$ to $4$, so that they give also a picture of the
convergence of our treatment in terms of the free modes of the
system. Notice that from eqs.~\eqref{force} and~\eqref{energy}, the
same results may be scaled to get the total power pumped by the
barrier, and are directly associated to the excess of mass, and of
kinetic energy through the relationship~\eqref{balance} imposed by the
steady state condition.  The results shows a clear difference between
the $C=0.2$ case, with very little dependence of $F_T$ on $\Gamma$,
and the high rate shift, $C=2.$, with a very rapid decay of the force
for increasing damping.

\section{Conclusions}
\label{conclusions}
Colloidal particles when subjected to external driving forces exhibit
may properties which are different from those of equilibrium
systems. In the present paper we have described the effect of a
barrier moving at constant velocity in a one dimensional colloidal
fluid in the approximation that the solvent is unaffected by the
barrier.  In contrast with previous approaches which have considered
only overdamped dynamics, we have studied the case where inertia plays
a role. The two major effects of inertial terms are first to determine
the appearance of a wake structure, completely absent in the DDF
treatment and of an infinite set of characteristic lengths in the
regions near the moving barrier; and second to produce not only a
strongly structured density distribution near the barrier, and the
associated current density, but also the higher order moments of the
velocity distribution distribution, which may be represented as a
local temperature profile, very different from that of the
thermalizing bath, and which shows interesting characteristics. It is
also interesting that the method used here, based on the natural
expansion of the distribution $P(X,V,\tau)$ for free systems, gives an
intuitive connection with the previous results based on the DDF
treatment, i.e. using the density distribution $\rho(X,\tau)=\rho_o
\Phi_0(X,\tau)$ as the only relevant field.  That approach is
recovered in as the limit of the simplest description of $P(X,V,\tau)$
in terms of the first free mode (the only one with a purely diffusive
dynamics, without an exponential decay time). The local balance of
mass and force reproduce the DDF result of a Smoluchowski equation.
To achieve the equivalent local balances for the energy, heat
currents, etc..., we have to enlarge the parametrization for
$P(X,V,\tau)$, to include exponential decaying modes, which represent
the effects of the inertial dynamics of the particles.

We have analyzed here only the simplest case, of one-dimensional
spatial distributions in the dilute, ideal gas, limit. The equivalent
results under other geometrical conditions, when the particles can
bypass the moving barrier~\cite{3d,Dzubiella}, and including the
effects of the particle interactions~\cite{pennatarazona,3d}, have
been explored under the DDF assumptions, and it would be interesting
to generalize them to the present approach.

It is perhaps worth to comment that
the wake region is not specific to the flow of particles with
inertia.
Very similar effects were also found in higher dimensions for overdamped, 
Brownian particles
driven past colloids, which act in this case as the potential barriers
\cite{Dzubiella}. 
As shown 
by Penna et al. \cite{Pennadzu} there exists a sum rule stating
that the integral for the wake in any transversal plane to the
direction of the drift vanishes, so that the
depletion along the axis through the obstacle 
is exactly cancelled by the contribution from the lateral wings.  
Such a sum
rule is valid in any dimension, but of course, in $D=1$ implies that there
is no wake at all.
The presence of a wake structure in $D=1$, in the inertial case, 
should correspond to a 
breaking of the sum rule for its transverse integral in $D>1$.

The present results perhaps are of relevance for
microfluidic devices where colloidal particles move along narrow
channels in order to understand what external forces are needed to
induce a drift in the presence of Brownian fluctuations. The
hydrodynamics interaction, which have been neglected here, could not
be important in one dimension due to the screening effect, and we may
expect that, at least at a qualitative level, the predictions made
here could be accessible to experimental observation.

P.T. acknowledges financial support from the Ministerio de Educacion y
Ciencia of Spain, under grant FIS2007-65869-C03-01, and the Comunidad
Autonoma de Madrid, under grant 0505/ESP-0299.

\newpage
\section*{Caption List}

{\bf Caption Fig. 1} 

{Steady state scaled density profile,
$\Phi_0(X)=\rho(X)/\rho_o$,  in the reference frame of the moving  parabolic 
barrier.
The barrier strength is $\kappa=10$, its width is $2$ and
moves at a relatively low velocity, $C=0.2$, while the 
damping constant $\Gamma$ takes on several values.  
Panel (b) shows the structure of 
the region within the barrier ($-1\leq X\leq 1$), and the inertial 
wake left behind by the advancing barrier. The position $X$ is relative
to the barrier. Adimensional units \eqref{adim1}-\eqref{adim2} are used for
all the quantities.}

{\bf Caption Fig. 2} 

{Scaled density profile, $\Phi_0(X)=\rho(X)/\rho_o$, 
at the front (a) and wake (b) regions in 
Fig.(1) is presented in reduced distances $(X \pm 1) C \Gamma$,
to take into account the natural decay length of the zeroth order mode.
The front structure curves collapse for  $\Gamma \geq 1$, while the
wake region is reduced for increasing $\Gamma$. The position $X$ is relative
to the barrier. Adimensional units \eqref{adim1}-\eqref{adim2} are used for
all the quantities.}

{\bf Caption Fig. 3}

\begin{figure}[htb]
\includegraphics[clip=true,width=120mm,keepaspectratio]{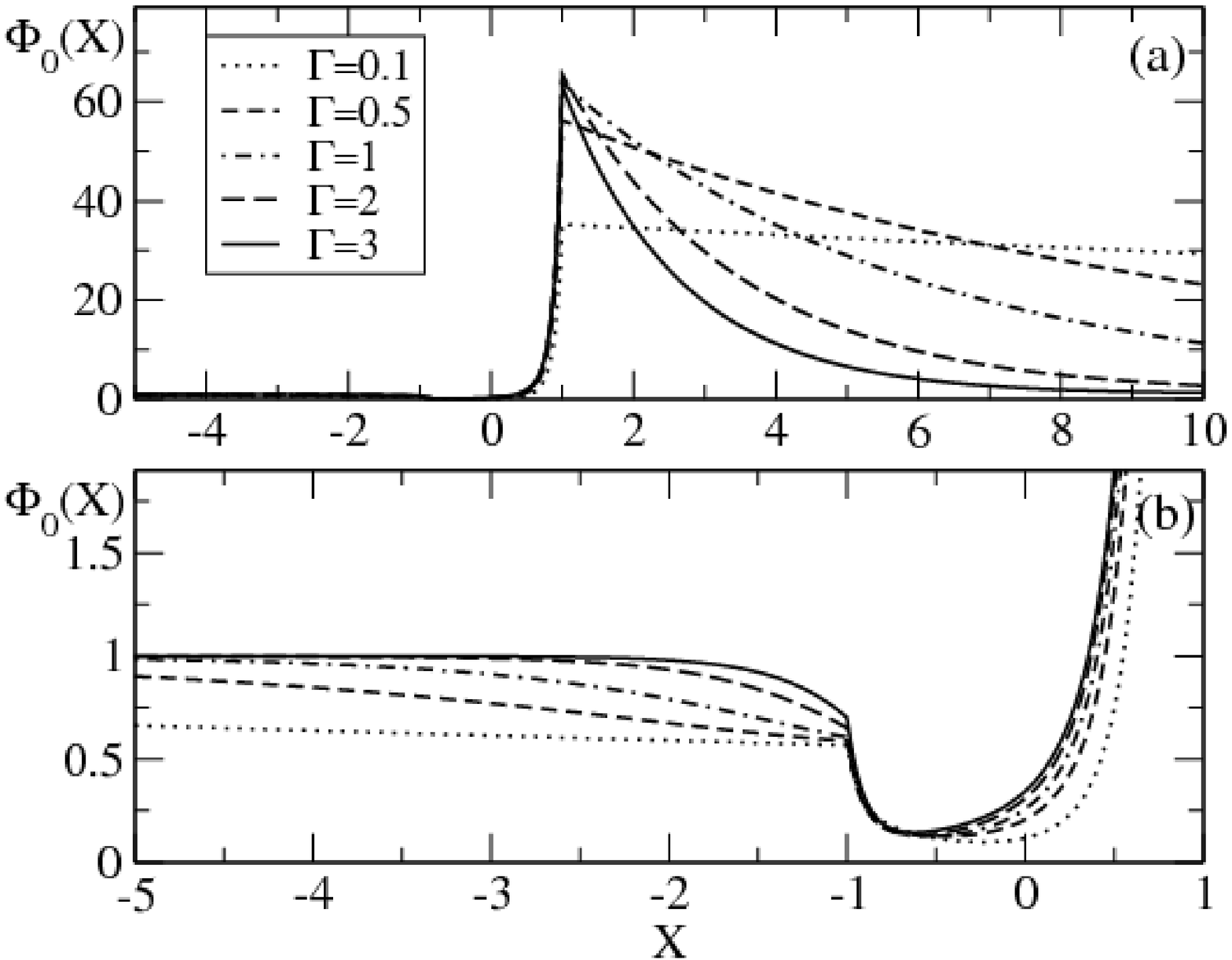}
\label{fig:1}
\end{figure}

\begin{figure}[htb]
\includegraphics[clip=true,width=120mm,keepaspectratio]{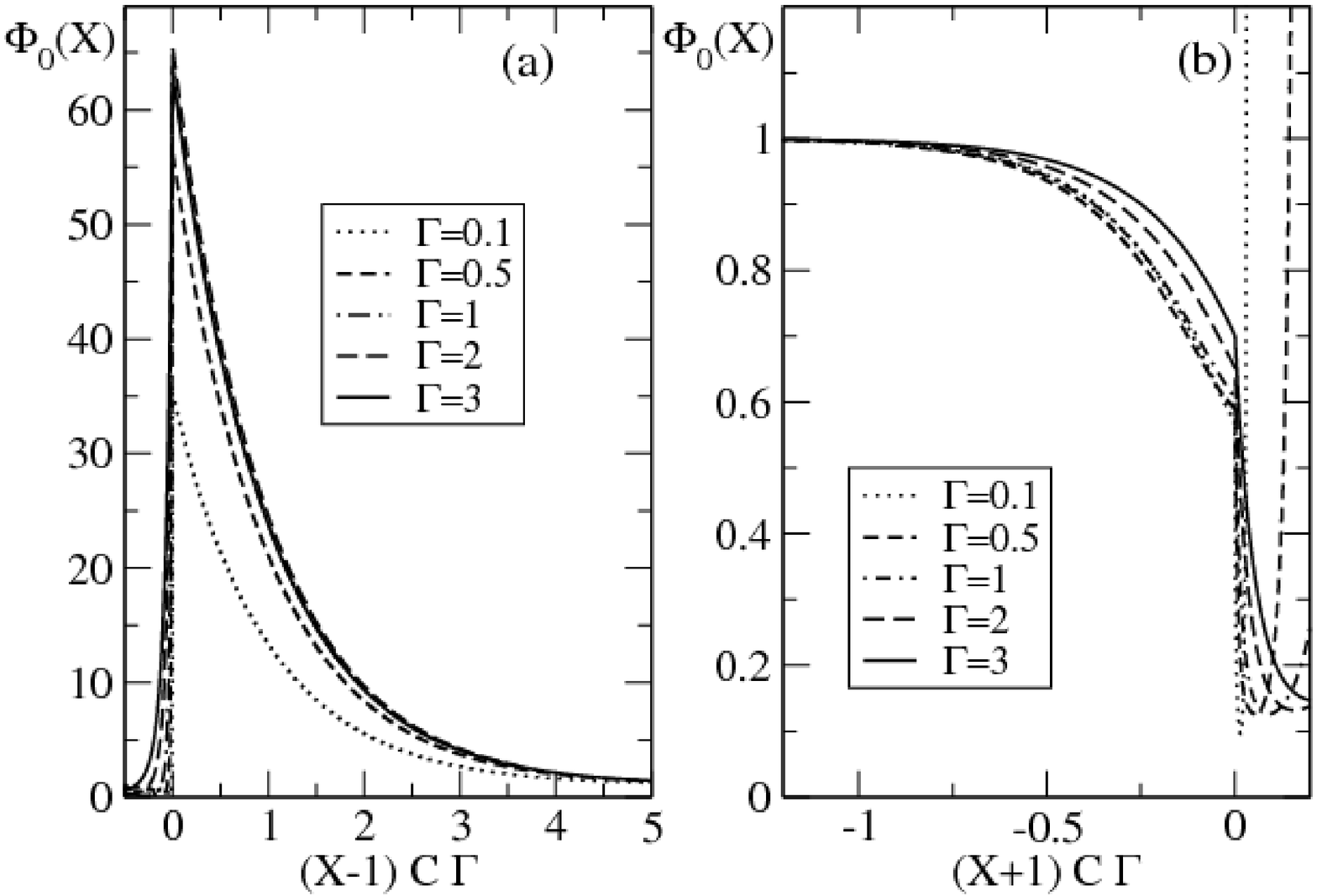}

\label{fig:2}
\end{figure}

\begin{figure}[htb]
\includegraphics[clip=true,width=120mm,keepaspectratio]{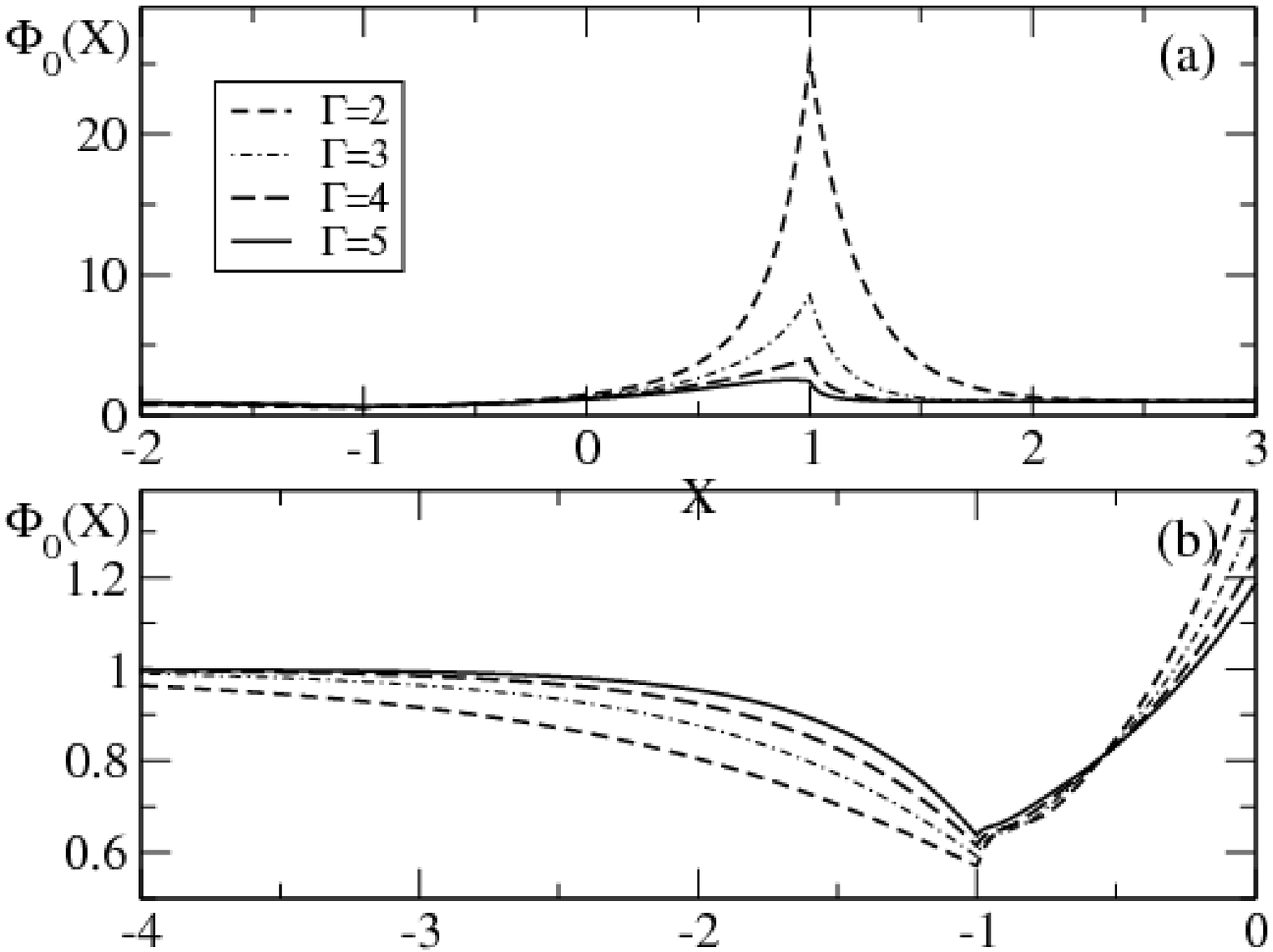}

\label{fig:3}
\end{figure}

\begin{figure}[htb]
\includegraphics[clip=true,width=120mm,keepaspectratio]{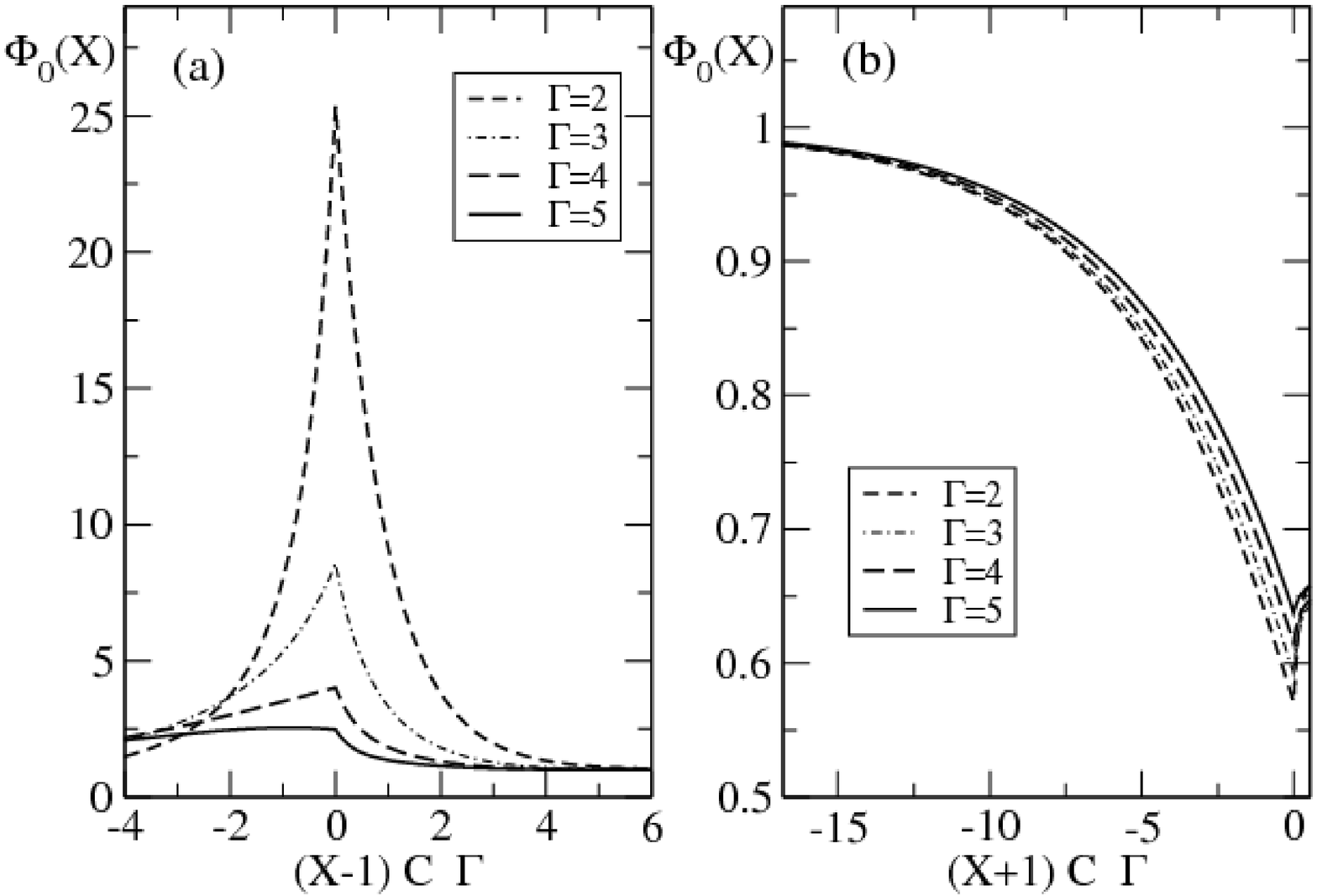}

\label{fig:4}
\end{figure}

\begin{figure}[htb]
\includegraphics[clip=true,width=120mm,keepaspectratio]{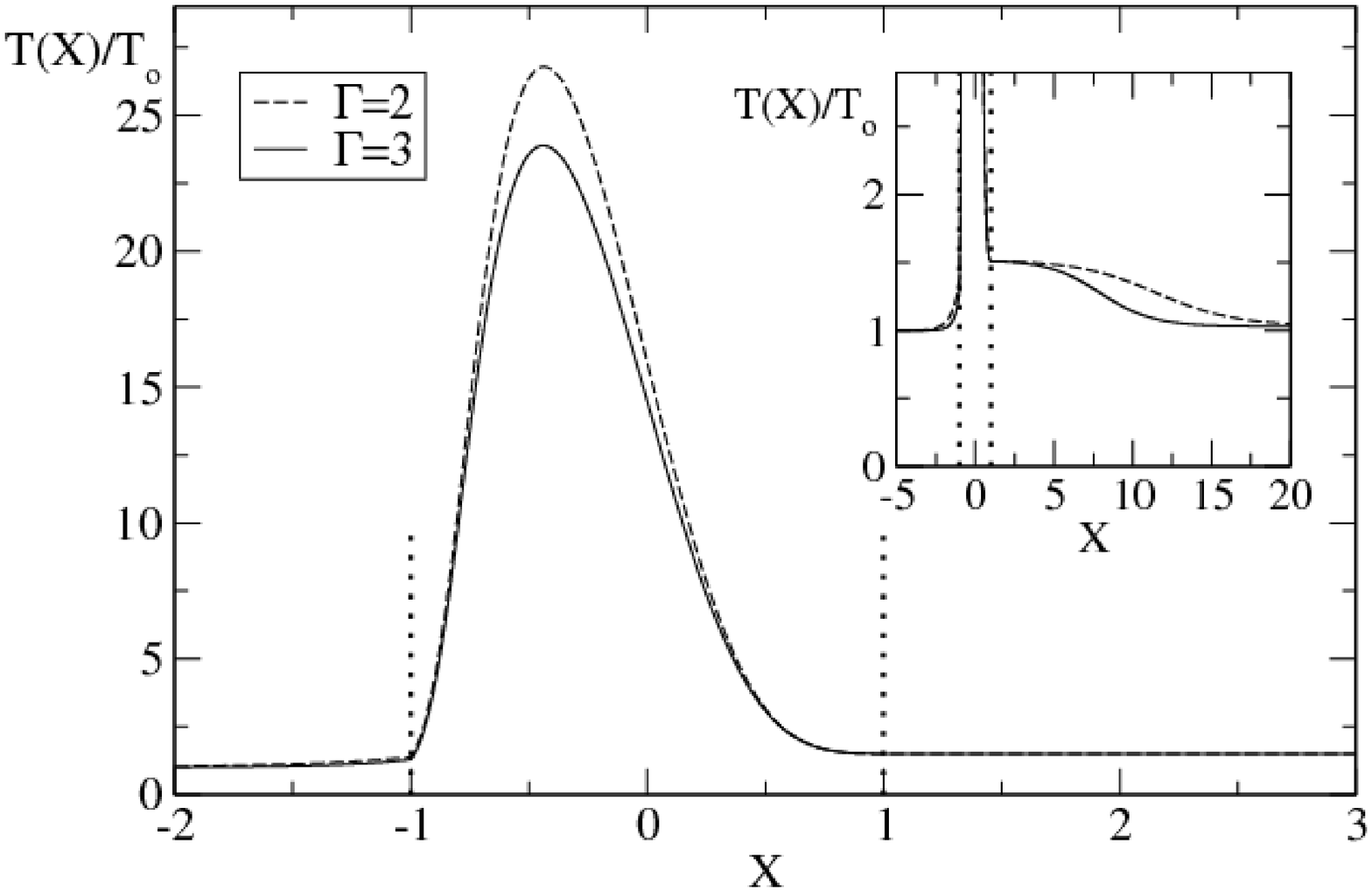}
\label{fig:5}
\end{figure}

{Steady state density distribution, $\Phi_0(X)=\rho(X)/\rho_o$, 
induced by 
the same potential barrier as in Fig. \ref{fig:1}, but moving at a higher
velocity, $C=2.$ The damping constants 
are $\Gamma=5$ (full line),  $\Gamma=4$ (long dashed line),
$\Gamma=3$ (dot-dashed line), $\Gamma=2$ (short dashed line).
Panel (a) gives a general view of the high
density structure at the advancing front. Panel (b) shows the structure of
the depleted region within the barrier ($-1\leq X\leq 1$), and the inertial
wake leaved behind by the advancing barrier.
The position $X$ is relative
to the barrier. Adimensional units \eqref{adim1}-\eqref{adim2} are used for
all the quantities.} 

{\bf Caption Fig. 4}

{The structure of the relative density $\Phi_0(X)=\rho(X)/\rho_o$, 
at the front (a) and wake (b) regions in
Fig.(3) is presented in reduced distances $(X \pm 1) C \Gamma$,
to take into account the natural decay length of the zeroth order mode.
The position $X$ is relative
to the barrier. Adimensional units \eqref{adim1}-\eqref{adim2} are used for
all the quantities.} 

{\bf Caption Fig. 5}

{Steady state temperature profile 
induced by a parabolic potential 
barrier shifted at rate  $C=0.2$.
The position $X$ is relative to the barrier and the vertical dotted lines are 
the barrier edges. 
The inset shows the structure of for $T(X)\approx T_o$.
Adimensional units \eqref{adim1}-\eqref{adim2} are used 
for all the quantities.} 

{\bf Caption Fig. 6}

{Steady state temperature profile induced  
by a rapid  drift,  
$C=2.$, of the parabolic potential barrier.  The position $X$ is relative 
to the barrier and the vertical dotted lines are the barrier edges.
Adimensional units \eqref{adim1}-\eqref{adim2} are used for
all the quantities.}

{\bf Caption Fig. 7} 
 
{Steady state total force 
exerted by the moving barrier on the particles
for the shifting rates $C=0.2$ and $C=2.$ The full lines are obtained using
the expansion \eqref{godf} in 
terms of the free modes up to $\mu_{max}=4$, the
dashed lines up to $\mu_{max}=2$ and the dotted lines up to
$\mu_{max}=2$. Adimensional units eqs.\eqref{adim1}-\eqref{adim2} are used
for all the quantities.}

\newpage
{\bf Fig. 1}
\vspace{3cm} 

\newpage
{\bf Fig. 2}
\vspace{3cm} 

\newpage
{\bf Fig. 3} 
\vspace{4cm}
\newpage
{\bf Fig. 4} 
\vspace{3cm}
\newpage
{\bf Fig. 5} 
\vspace{3cm}
\newpage
{\bf Fig. 6} 
\vspace{3cm} 
\begin{figure}[htb]
\includegraphics[clip=true,width=120mm,keepaspectratio]{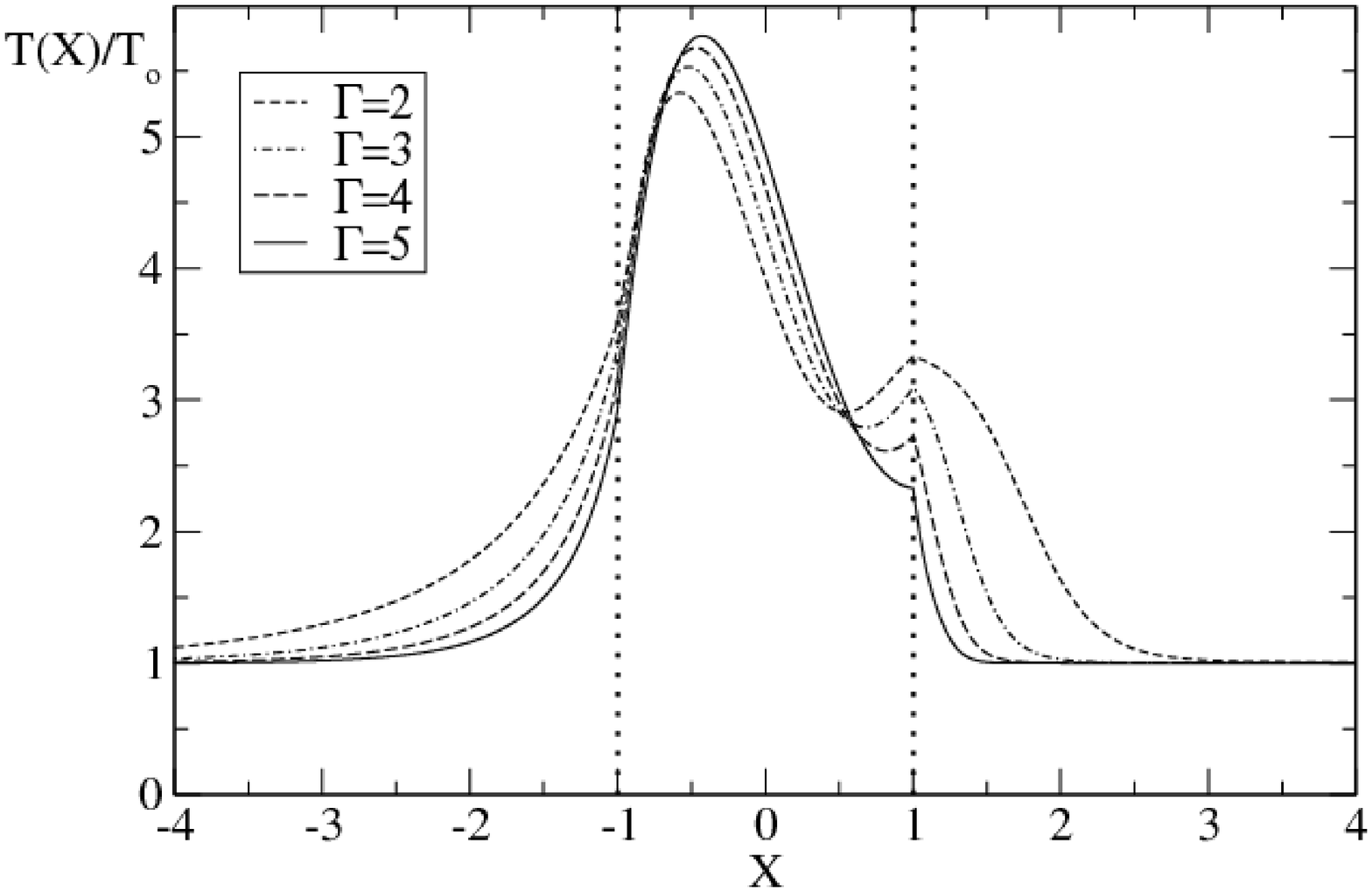}
\label{fig:6}
\end{figure}
\newpage
{\bf Fig. 7}
\vspace{3cm} 
\begin{figure}[htb]
\includegraphics[clip=true,width=120mm,keepaspectratio]{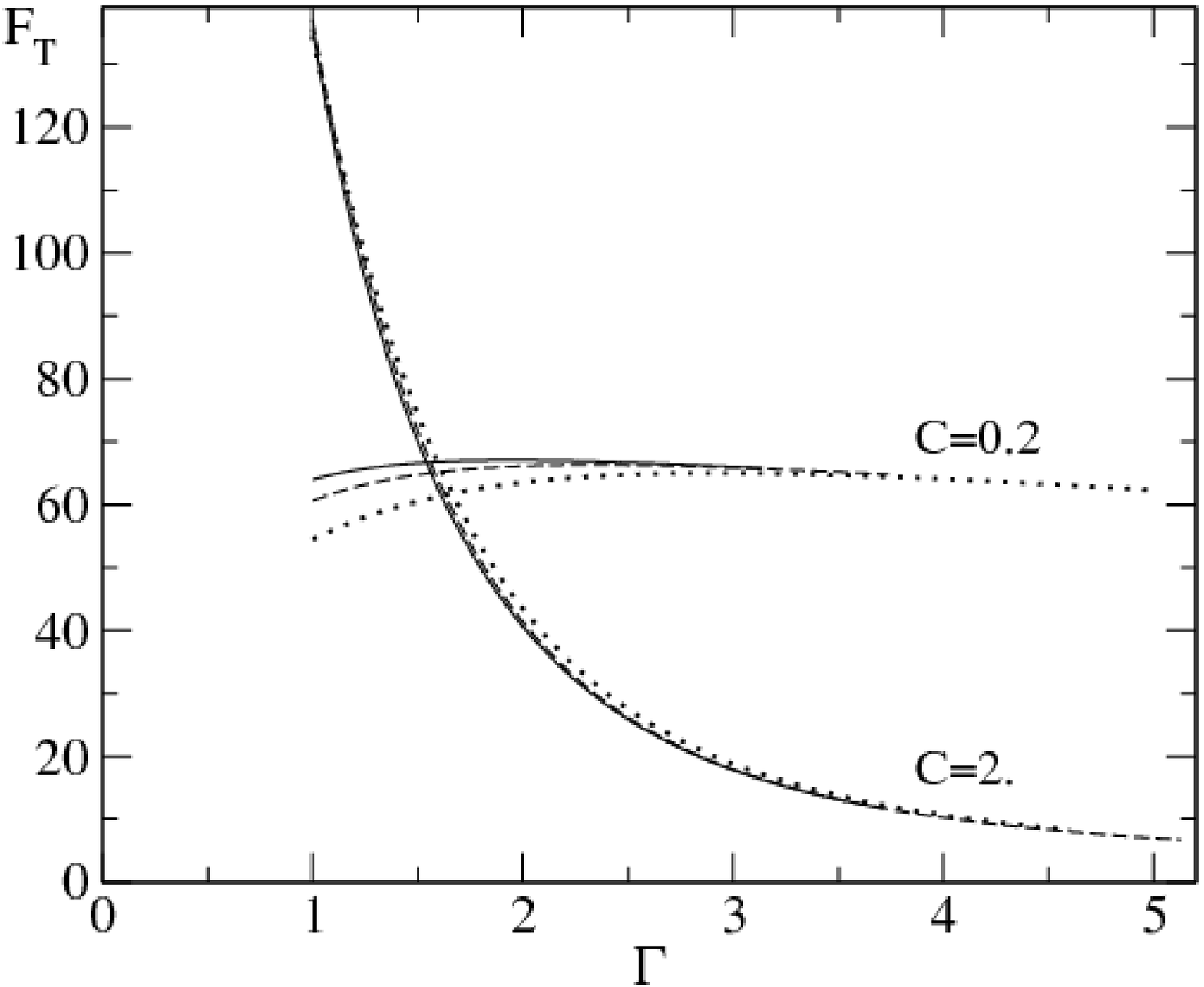}
\label{fig:7}
\end{figure}

\end{document}